\documentclass[journal]{journal}
\newcommand{\indep}{\perp \!\!\! \perp} 

\ifCLASSINFOpdf
 \usepackage[pdftex]{graphicx}
\else
 \usepackage[dvips]{graphicx}
\fi
\let\labelindent\relax
\usepackage{enumitem}
\usepackage{hyperref}

\usepackage[cmex10]{amsmath}
\usepackage{amsmath}

\begin{document}
%
\title{Hierarchical Bayesian Claim Count modeling with Overdispersed Outcome and Mismeasured Covariates in Actuarial Practice}
%
%
%

\author{
Minkun~Kim,~
Martin~Crane,~
Marija~Bezbradica,~\IEEEmembership{ADAPT~Centre,~Ireland}
\thanks{Minkun Kim is with ADAPT Centre, Dublin City University \\
58 Albert College Park, Whitehall, D09PX21, Dublin, Ireland \\ e-mail: minkun.kim4@mail.dcu.ie.}
}

\maketitle
%
%

\markboth{Journal of \LaTeX\ Class Files,~Vol.~6, No.~1, January~2007}%
{Shell \MakeLowercase{\textit{et al.}}: Bare Demo of IEEEtran.cls for Journals}
%



\thispagestyle{empty}

\begin{abstract}
The problem of overdispersed claim counts and mismeasured covariates is common in insurance. On the one hand, the presence of overdispersion in the count data violates the homogeneity assumption, and on the other hand, measurement errors in covariates highlight the model risk issue in actuarial practice. The consequence can be inaccurate premium pricing which would negatively affect business competitiveness. Our goal is to address these two modelling problems simultaneously by capturing the unobservable correlations between observations that arise from overdispersed outcome and mismeasured covariate in actuarial process. To this end, we establish novel connections between the count-based generalized linear mixed model (GLMM) and a popular error-correction tool for non-linear modelling - Simulation Extrapolation (SIMEX). We consider a modelling framework based on the hierarchical Bayesian paradigm. To our knowledge, the approach of combining a hierarchical Bayes with SIMEX has not previously been discussed in the literature. We demonstrate the applicability of our approach on the workplace absenteeism data. Our results indicate that the hierarchical Bayesian GLMM incorporated with the SIMEX outperforms naïve GLMM / SIMEX in terms of goodness of fit.

\end{abstract}

\begin{IEEEkeywords}
Count-based regression, Hierarchical Bayes, Negative Binomial GLMM, SIMEX, overdispersion, measurement error.
\end{IEEEkeywords}

%
\IEEEpeerreviewmaketitle

\section{Introduction}
%
%
%
%
\IEEEPARstart{A}{s} claims from policyholders are directly translated into business losses, it is imperative for insurers to properly estimate total claim amounts. The aggregate claim amounts up to time \(t\) received by an insurance company can be defined as \(S(t) = \sum_{i=1}^{N(t)} claim_{i} = claim_{1} + claim_{2} + \ldots + claim_{N(t)}\) in which \(claim_{i}\) is the amount of a claim in \(i\)-\(th\) case, and \(N(t)\) is total claim count up to time \(t\). In order to successfully evaluate \(S(t)\), the first step is to model the aggregate claim count \(N(t)\) by assuming it follows some process such as a Poisson process (see Cossette et al.~\cite{cossette2020ruin} for further details). However, as Payne et al. point out, in practice, unobservable differences between individual claim count records often result in overdispersion\footnote{Overdispersion refers to high variability in data, much larger than the data mean. It is assumed that some unobserved correlation can produce additional dispersion in the observed counts, which results from data heterogeneity~\cite{winkelmann2008econometric}.}, violating the Poisson homogeneity assumption~\cite{payne2017approaches}. In addition, measurement errors in covariates are not rare due to the complex structure of the information-intensive actuarial task, which triggers ``Model risk"\footnote{`Model risk' in actuarial practice refers to a risk of adverse consequences from decisions based on incorrect data or wrong models~\cite{black2018model}.} in the insurance context~\cite{glowalla2013managing}. A consequence of this confluence between overdispersed count outcomes and mismeasured covariates is an inaccurate premium pricing model built on the unfair risk distributions for each insured class of policyholders\footnote{This is because ``within-group variation" deteriorates with noise and ANOVA tests fail to capture relevant variation due to unobserved ``within-group heterogeneity"~\cite{gormley2014common}.}, hence they can be wrongly charged at an incorrect premium rate. 

There is substantial literature on approaches to correct for overdispersion, as well as for measurement error. Focusing on Bayesian perspectives, we discuss some literature below.     
\subsubsection{Overdispersion:}
Wang highlights the generalized linear mixed model (GLMM) as a ``parametric mixture" of a Poisson likelihood and a Gamma prior, using the rule of a Bayesian conjugate relationship~\cite{wang2011one}. This mixture technique naturally induces a Negative Binomial GLMM which has an extra variation (dispersion) parameter to account for unidentified confounders that cause overdispersion. In complex settings, however, Lloyd-Smith shows that the Negative Binomial GLMM with standard maximum likelihood estimator (MLE) for the dispersion parameter might fail to make a proper inference when the model requires high-dimensional numerical integration (with multiple predictors) or when there are not enough samples~\cite{lloyd2007maximum}. In this respect, Zhou et al. add a ``hierarchical Bayes" on the top of the GLMM technique to estimate the dispersion parameter without using MLE~\cite{zhou2012lognormal}. The main reason why ``hierarchical Bayes" is preferred over the pure ``parametric mixture" is that while the ``parametric mixture" technique can only apply the unobserved variation parameter (such as the dispersion parameter) to each observation, ``hierarchical Bayes" can incorporate such variation with multiple levels of their aggregation. Built on the multiple hierarchical levels, the ``hierarchical Bayes" explores the nested structure in the parameter space to deal with heterogeneity at each level of aggregation. This naturally introduces correlation across entire observations in the data space so that the data heterogeneity is absorbed by this correlation~\cite{shi2012bayesian}.

\subsubsection{Measurement error}
When it comes to the mismeasured covariates in Count data modelling, \emph{Simulation and Extrapolation} (SIMEX) is a largely dominant approach discussed in the literature. The basic idea is that if the model discovers a measurement error trend - a relationship between the degrees of measurement error and the resultant estimation biases - then the true estimation can be obtained by extrapolating it back to the setting without measurement error. It has gained popularity because the exact distribution of the contaminated covariate does not need to be known~\cite{oh2018considerations}. In a Bayesian context, however, the distribution of the covariate should be postulated, thus the Bayesian generative principle dictates the introduction of prior knowledge to perform random sampling on that distribution. This approach is analogous to the solution of the ``inverse problem"~\footnote{The inverse problem refers to the situation where the outcome is given, but the condition that might have led to the outcome is unknown, thus sought~\cite{tarantola2006popper}}. The canonical paper by Tarantola~\cite{tarantola2006popper} provides useful insight on how the unclear data model behaviors can be estimated with the inverse simulation technique. In a nutshell, this technique selects the most likely parameter samples and then generates a variety of scenarios of true data behaviors by feeding those selected posterior samples into the contaminated data model. 

This paper is organized as follows. In Section 2, we overview the Hierarchical Bayes framework and some error correction tools such as the Simulation Extrapolation algorithm. We also discuss how hierarchical Bayesian regression is combined with this error-correction tool. Section 3 presents specifics on model selection criteria for this study. In Section 4, we validate our approach on both simulated and real data and demonstrate how unexplained correlations among observations that are associated with overdispersion and measurement error can be captured and addressed by this hybrid Bayesian method. Finally, our study is summarized and a discussion of this study is drawn in Section 5.

\section{The Model}
Our interest lies in simultaneously counteracting two problems: 1) overdispersed outcome and 2) mismeasured covariates. We here present hierarchical Bayesian GLMM coupled with SIMEX. We initially consider a \(p\)-dimensional Negative Binomial GLMM to model the overdispersed count-based outcome \(y\). For each individual record \(i = 1,..,n\) count data points, let \(X_{i}=[1,x^{1}_{i},x^{2}_{i},..x^{p}_{i}]^T\) be the vector of covariates subject to measurement error. Assuming overdispersion is a product of unknown variability between cohorts in \(y\), for each unknown cohort \(j = 1,..,m\), let \(\gamma_{j}\) be the random dispersion effect term of the underlying Negative Binomial distribution of \(y_{ij}\), expressed as \( y_{ij} \sim \mathbf{NB(}\eta_{ij}, \; \gamma_{j}\mathbf{)} \). As the count data \(y_{ij}\) is characterized by discreteness and non-negativity, each one of them is with subject-specific mean \(\eta_{ij}=\exp(X_{ij}^{T}\boldsymbol{\underline{\beta}_{j}}) \) with a coefficient vector \(\boldsymbol{\underline{\beta}_{j}}=(\beta^{0}_{j},\beta^{1}_{j},..\beta^{P}_{j})\), and group-specific random effect term \(C_{j} = \exp(\beta^{0}_{j})\) to address the data heterogeneity; i.e., 
\begin{equation}
    \eta_{ij} = C_{j} \cdot \exp(\beta_{j}^{1}x_{ij}^{1}+\beta_{j}^{2}x_{ij}^{2}+ \dots + \beta_{j}^{p}x_{ij}^{p})
\end{equation} 

We propose a hierarchical prior \(\theta_{j} = (\underline{\boldsymbol{\beta}_{j}}, C_{j}, \gamma_{j})\):
\begin{equation}
\begin{split}
    \boldsymbol{\underline{\beta}_{j}}\;|\; 
    \underline{\mu}_{j}, \Lambda_{j}
    &\sim \mathbf{MVN}_{p}(\underline{\mu}_{j}, \Lambda_{j}:g_{j}\Sigma_{0}) \\
    C_{j}\;|\;k_{j} &\sim \mathbf{Ga}(rate=k_{j}, \;  shape=k_{j}/E[C_{j}]) \\
    \gamma_{j}\;|\;a_{j} &\sim \mathbf{Ga}(rate=a_{j}, \; shape=a_{j}/E[\gamma_{j}])
\end{split}
\end{equation}
with 
\begin{equation}
\begin{split}
    \underline{\mu}_{j}\;|\;\hat{\beta}_{GLM} &\sim \mathbf{MVN}_{p}(\hat{\beta}_{GLM}, \; \Sigma_{0}) \\
    \Lambda_{j}:g_{j}\Sigma_{0}\;|\;\nu,\;\tau &\sim \mathbf{IW}(dof=\nu, \; scale=\tau\Sigma_{E}) \\
    k_{j}\;|\;s,\;t &\sim \mathbf{Ga}(shape=s, \; scale=t) \\
    a_{j}\;|\;u,\;v &\sim \mathbf{Ga}(shape=u, \; scale=v)
\end{split}
\end{equation} where \(\underline{\mu}_{j}, \Lambda_{j}, k_{j}, a_{j}\) are the unknown hyper-prior parameters, \(\hat{\beta}_{GLM}\) stands for the initial estimate results of the Negative Binomial GLM, \(\Lambda_{j}: g_{j}\Sigma_{0}\) is \(g_{j}\hat{\sigma}^2(X^TX)^{-1}\) with \(g=nm\) to make \(\frac{g}{g+1}\) close to \(1\) so that \(\underline{\beta}_{j}\) is centered around \(0\)~\cite{albert2007bayesian}. \(\Sigma_{E}\) is expert's chosen variance matrix, and \(\nu, \tau, s, t, u, v\) are hyperparameters. In short, our hierarchical Bayesian GLMM setting with three different levels can be described as below:
\begin{itemize}
    \item data point level: \(y_{1j},y_{2j},\dots|\; \underline{X}_{j},\theta_{j}\)
    \item main parameter level: \(\theta_{j} = ( {\boldsymbol{\underline{\beta}_{j}},C_{j},\gamma_{j}\;|\;\delta_{j}} )\)
    \item hyper-prior parameter level: \(\delta_{j} = ( {\underline{\mu}_{j},\Lambda_{j},k_{j},a_{j}} )\)
\end{itemize} and the graphical representation of this scenario is shown in Figure (1).  
\begin{figure}[!t]
\centering
\includegraphics[width=2.5in]{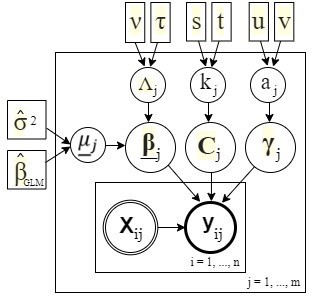}
\caption{Proposed directed acyclic graph (DAG) representation of the hierarchical Bayesian GLMM with the count-based outcome. Note that \(y_{ij}\) is overdispersed and \(X_{ij}\) is mismeasured.}
\label{fig_sim}
\end{figure} We expect that the model above can take into account unobserved confounders that cause overdispersion between cohorts in \(y_{ij}\), using two additional random effect terms - \(C_{j}, \gamma_{j}\) to examine the information shared across groups. However, this model is still naive due to the false information fed by the mismeasured covariates \(X_{ij}\). \\   

To make this model realistic, we can correct the \(\eta_{ij}\) in Equation (1) by introducing SIMEX that formulates the new distribution for the true covariates via extrapolation. For the sake of simplicity, suppose that we have a single covariate mismeasured. Our mismeasured covariate \(x_{i}\) is defined as \(x_{i} = U_{i}+\epsilon_{i}\) where \(U_{i}\) is the unobservable true covariate that we seek, and the measurement error follows \(\epsilon_{i} \sim \mathbf{N}(0, \sigma_{\epsilon}^2)\). If the covariate follows a Gaussian density, the distribution of the mismeasured covariate we have can be expressed as:
\begin{equation}
\begin{split}
     x_{i} &\sim \mathbf{N(}E[U_{i}], \; \sigma_{U}^2 + \sigma_{\epsilon}^2\mathbf{)} 
\end{split}
\end{equation} where \(\sigma_{U}^2 \indep \sigma_{\epsilon}^2\). In order to develop a tool to control over the variance of the target covariate distribution in Equation (4), assuming the variance of the measurement error \(\sigma_{\epsilon}^2\) is known, we can generate additional customized error term \(\sqrt{\lambda_{t}}\epsilon_{i}, \sim \mathbf{N(}0, \; \lambda_{t}\sigma_{\epsilon}^2\mathbf{)}\) where \(\lambda_{t}\) is a set of values to indicate the contamination level \(0=\lambda_{1} < \lambda_{2}, \dots < \lambda_{t}, \dots < \lambda_{T} \). Then the SIMEX suggests
the new subject-specific mean \(\eta_{ij}^{*}=\exp(W_{ij}(\lambda_{t})^{T}\boldsymbol{\underline{\beta}_{j}}) \) and Equation (1) can be re-defined as:
\begin{equation}
    \eta_{ij}^{*} = C_{j} \cdot \exp(\beta_{j}^{1}x_{ij}^{1} +\dots \beta_{j}^{*}w_{ij}^{*}(\lambda_{t}) + \dots \beta_{j}^{p}x_{ij}^{p})
\end{equation} where the new customized covariate \(w_{ij}^{*}(\lambda_{t}) = x_{ij} + \sqrt{\lambda_{t}} \epsilon_{i}\) with the value of \(\lambda_{t}\) chosen by researchers. Considering Equations (4) and (5), we can finally obtain the ultimate form of the distribution to simulate a new covariate under control as below:
\begin{equation}
\begin{split}
     w_{ij}^{*}(\lambda_{t}) &\sim \mathbf{N(}E[U_{ij}], \; \sigma_{U}^2 + (1+\lambda_{t})\sigma_{\epsilon}^2\mathbf{)}
\end{split}
\end{equation} in which the total variance increases as the value \(\lambda_{t}\) increases. Intuitively speaking, the extrapolation to find the parameters for the clean covariate can be made in the plane with the contamination level \(1+\lambda_{t}\) on the horizontal axis and a set of all resulting parameters \(\theta\) estimations on the vertical axis. Once their relationship is properly mapped and modelled, then the distribution for the clean covariate and a set of noise-free parameter values can be obtained by the quadratic extrapolation, substituting \(-1\) into \(\lambda_{t}\) which is the ideal case of no-measurement error~\cite{carroll2006measurement}. In our paper, we use this SIMEX extrapolation result \(\hat{\underline{\beta}}_{GLM}\) as initialization values to get the hyper-prior parameters \( \underline{\mu}_{j}\), which is briefly illustrated in Figure (2).    
\begin{figure}[!t]
\centering
\includegraphics[width=2.5in]{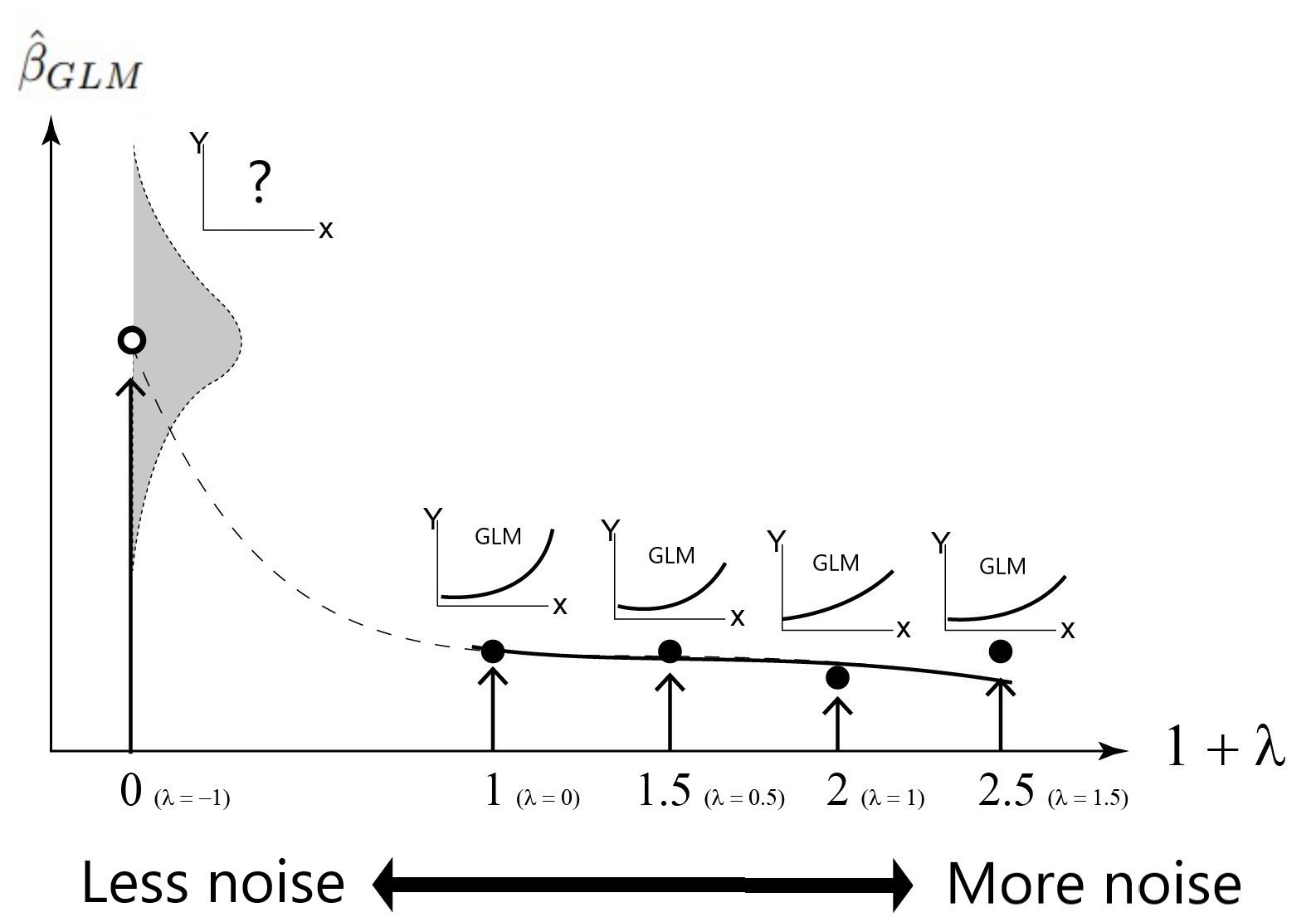}
\caption{A generic diagram of a SIMEX extrapolation to explain the impact of the data contamination level \((1+\lambda)\sigma_{\epsilon}^2\) on the estimates \(\theta\).}
\label{fig_sim}
\end{figure}

When it comes to the posterior inference, substituting \(\eta_{ij}^{*}\) in Equation (5) 
back into the outcome model \(\mathbf{NB}(\eta_{i,j}^{*}, \gamma_{j})\), we can update the main parameter \(\theta_{j}^{*} = (\underline{\boldsymbol{\beta}}_{j}^{*}, C_{j}^{*}, \gamma_{j}^{*})\) with typical Markov Chain Monte Carlo (MCMC) sampling. Each completion of a SIMEX simulation step provides a new covariate \(W_{ij}(\lambda_{t})^{T}\) which passes into the joint density functions to approximate the posterior samples \( \theta_{j}^{*}\). For example, in MCMC, both the main parameters \(\theta_{j} = ( {\boldsymbol{\underline{\beta}}_{j},C_{j},\gamma_{j}\;|\;\delta_{j}} )\), and the hyper-prior parameters \(\delta_{j} = ( {\underline{\mu}_{j},\Lambda_{j},k_{j},a_{j}} )\) can be iteratively sampled together from their conditional joint densities built upon the hierarchical Bayes structure as described below. Let \(\theta^{(h)}\)denote the state of parameter \(\theta\) in the \(h\) iteration, for h=1,...,H, and let the intercepts \(C_{j}\) and slopes \(\boldsymbol{\underline{\beta}}_{j}\) are allowed to vary by each cohort. 
\begin{enumerate}[label={\arabic*.}]
    \item Choose an initial configuration for each parameter, say \(\underline{\boldsymbol{\beta}}_{j}^{(0)}, \boldsymbol{C}_{j}^{(0)}, \boldsymbol{\gamma}_{j}^{(0)}, \underline{\mu}_{j}^{(0)}, \Lambda_{j}^{(0)}, k_{j}^{(0)}, a_{j}^{(0)}\), and some fixed values: \(\nu, \tau, s, t, u, v,\) and \(\lambda_{t}\). 

    \item Update the main parameters and hyper-prior parameters: \(\underline{\boldsymbol{\beta}}_{j}^{(h-1)}, \boldsymbol{C}_{j}^{(h-1)}, \boldsymbol{\gamma}_{j}^{(h-1)}, \underline{\mu}_{j}^{(h-1)}, \Lambda_{j}^{(h-1)}, k_{j}^{(h-1)}, a_{j}^{(h-1)}\) until convergence: 
        \begin{enumerate}[label={(\alph*)}]
            \item Calculate \(\hat{\beta}_{GLM}\) and \(\hat{\sigma}_{GLM}^2\) with SIMEX Extrapolation based on \(\underline{\boldsymbol{\beta}}_{j}^{(h-1)}, \boldsymbol{C}_{j}^{(h-1)}\)
        
            \item Sample \(\Lambda_{j}^{(h)}\) to get ``Varying slope" from: \\ \( \mathbf{IW}_{P} \Big( \Lambda_{j}\Big|\; \nu+P, \\
            \tau\Sigma_{E}+\sum_{p=1}^{P}(\beta_{jp}^{(h-1)} - \mu_{jp}^{(h-1)})(\beta_{jp}^{(h-1)} - \mu_{jp}^{(h-1)})^{T} \Big)\)
            
            \item Sample \( \underline{\mu}_{j}^{(h)}\) to get ``Varying slope" from: \\ \( \begin{aligned} \mathbf{MVN}_{p} \Big( \underline{\mu}_{j}\Big|\; \frac{(nm)^{-1}}{(nm)^{-1}+1}\hat{\beta}_{GLM} + \frac{1}{(nm)^{-1}+1}\boldsymbol{\underline{\beta}_{j}}^{(h-1)}, \\ \frac{1}{(nm)^{-1}+1}\Lambda_{j}^{(h)}\Big)
            \end{aligned}\) 
            
            \item Sample \(\underline{\boldsymbol{\beta}}_{j}^{(h)}\) error-corrected ``Varying slope" from: \\ \( \begin{aligned} \prod_{i=1}^{n}&\mathbf{NB}\Big( y_{i}\;\Big|\; e^{ W_{i}(\lambda_{t})^{T}\boldsymbol{\underline{\beta}_{j}}^{(h-1)} }, \; \gamma_{j}^{(h-1)}\Big) \times \\ &\mathbf{MVN}_{p}\Big( \boldsymbol{\underline{\beta}_{j}} \;\Big|\;\underline{\mu}_{j}^{(h)}, \; \Lambda_{j}^{h}\Big) 
            \end{aligned} \)
            
            \item Sample \( k_{j}^{(h)}\) to get ``Varying Intercept" from: \\ \( \mathbf{Ga}\Big( k_{j} \;\Big|\; s, \; t \Big) \) 
            
            \item Sample \( a_{j}^{(h)}\) to get ``Dispersion"  from: \\ \(\mathbf{Ga}\Big( a_{j} \;\Big|\; k_{j}^{(h)} + u, \; \gamma_{j}^{(h-1)}+v \Big) \)
            
            \item Sample \(C_{j}^{(h)}\) error-corrected ``Varying intercept" from: \\ \(\begin{aligned} \prod_{i=1}^{n}&\mathbf{NB}\Big( y_{i}\;\Big|\;e^{ W_{i}(\lambda_{t})^{T}\boldsymbol{\underline{\beta}_{j}}^{(h)} }, \; \gamma_{j}^{(h-1)}\Big) \times \\ &\mathbf{Ga}\Big( C_{j} \;\Big|\;k_{j}^{(h)}, \; k_{j}h({\sum_{z=1}^{h}C_{j}^{z}})^{-1} \Big)  
            \end{aligned} \)
            
            \item Sample \(\gamma_{j}^{(h)}\) error-corrected ``Dispersion" from: \\ \(\begin{aligned} \prod_{i=1}^{n}&\mathbf{NB}\Big( y_{i}\;\Big|\;e^{ W_{i}(\lambda_{t})^{T}\boldsymbol{\underline{\beta}_{j}}^{(h)} }, \; \gamma_{j}^{(h)}\Big) \times \\ &\mathbf{Ga}\Big( \gamma_{j} \;\Big|\;a_{j}^{(h)}, \; a_{j}h({\sum_{z=1}^{h}C_{j}^{z}})^{-1} \Big)  
            \end{aligned}   \)
         
        \end{enumerate}
    \item Select the posterior samples \(\theta_{j}^{*} = (\underline{\boldsymbol{\beta}}^{*}, C_{j}^{*}, \gamma_{j}^{*})\) via the acceptance rule with the joint density: 
        \begin{itemize}
            \item \(\pi(\theta_{j}) = \prod_{i=1}^{n}\mathbf{L}(\theta_{j}; y_{ij},X_{ij},w_{ij})\cdot \pi_{0}(\theta_{j})\)
        \end{itemize}

    \item If \(P_{accept} = \frac{\pi(\theta_{j}^{(h)})}{\pi(\theta_{j}^{(h-1)})} \geq 1\), then \\ \( \boldsymbol{\underline{\beta}_{j}}^{(h)}= \boldsymbol{\underline{\beta}_{j}}^{\textrm{*}}, \; \boldsymbol{C_{j}}^{(h)} = \boldsymbol{C_{j}}^{\textrm{*}}, \; \boldsymbol{\gamma_{j}}^{(h)} = \boldsymbol{\gamma_{j}}^{\textrm{*}} \) \\

    \item Cycle until achieving convergence.
\end{enumerate} 

\section{Model Selection Tool}
We consider a metric that accounts for both model fit and overfitting. The literature has largely focused on information criterion-based fit indices as a means of choosing the simplest model that best fits the given data. This includes selecting a model with the smallest Akaike information criterion (AIC) value. However, AIC typically does not work well because the nonlinear relationship in the GLMM violates the assumptions of AIC~\cite{anderson2004model}. In AIC, all main parameters should follow a roughly Multivariate Gaussian density, but it does not make the case in GLMM. For example, \(C_{j}, \gamma_{j}\) follows a Gamma in our study. In this regard, we consider an alternative - Watanabe Akaike Information Criteria (WAIC) - to the AIC. The WAIC is a Bayesian extension to the AIC, and it eliminates the distributional constraints such as the normality of residuals, etc. This is because WAIC uses entire posterior samples without any distributional assumption. The form of the WAIC is ``\(-2\cdot[\)LPPD \(-\) penalty\(]\)", and just like AIC, it is based on \emph{Log Point-wise Probability Density} (LPPD). 
\begin{equation*}
    \begin{aligned}
    &\mathbf{WAIC}(y_{i},X_{i}, \theta^{*}) \\ &=-2 \cdot \Big[ \mathbf{LPPD} - \sum_{i=1}^{N} VAR ( \log(\frac{1}{J}\sum_{j=1}^{J}\mathbf{L}(\theta_{j}^{*}; y_{i}, X_{i})) \Big]
    \end{aligned}
\end{equation*} 
with 
\begin{equation*}
    \mathbf{LPPD}(y_{i},X_{i}, \theta^{*}) = \sum_{i=1}^{N}\log( \frac{1}{J}\sum_{j=1}^{J}\mathbf{L}(\theta_{j}^{*}; y_{i},X_{i}) )
\end{equation*} In WAIC, its special penalty term is proportional to the variance of the predictive distribution, and each observation can have its own penalty value. Note that by identifying the points \(y_{i}\) that are sensitive to the certain posterior samples estimation, the WAIC allows for assessing overfitting at the level of individual observation~\cite{gelman2014understanding}.

\section{Numerical Example}
We consider a real data example to demonstrate how this model can be used for claim count prediction, given that the outcome variable is overdispersed and the covariates are mismeasured. The data set contains 36 employees' absence hour count information on \(n\)=740 records collected by a courier company from July 2007 to July 2010 in Brazil. Each entry comes with 20 additional attributes \(X: x^{P=1\sim20}\). The time values are rounded to the nearest integer in order to discretize the continuous time variable that measures the absence hour per case taken by each employee. The absence hour count information is possibly connected to the future benefit claims as discussed in Section 1.

\subsection{Exploratory Data Analysis}
A histogram of these absence hours per case \(y_{ij}\) is shown in Figure (3) which reveals the highly skewed distribution. We can see the majority of the records are not longer than 8 hours. The mean and the variance of the absence hours per case are 6.92 and 177.7 respectively, which indicates the presence of substantial overdispersion on the target variable \(y\). Checking the multicollinearity and the statistical significance, we have decided to focus on the five covariates - 1) Employee ID:\(x^{P=1}\), 2) BMI:\(x^{P=2}\), 3) Service Time:\(x^{P=3}\), 4) Distance-to-work:\(x^{P=4}\), and 5) Education:\(x^{P=5}\). Given that the variable ``Distance-to-work":\(x^{P=4}\) is the error-prone predictor, as discussed in Section 2, the measurement error variance \(\sigma_{\epsilon^{4}}^2\) associated with the covariate \(x^{4}\) should be estimated. Carroll et al suggest that if some independent replicate measurements of the same covariate are available, the unknown error variance \(\sigma_{\epsilon}^2\) can be estimated~\cite{carroll2006measurement}. Let the observed mismeasured covariate \(x_{i}^{P=4} = u_{i}^{P=4} + \epsilon_{i}\) where \(u_{i}\) is the unobservable true covariate. Now when we introduce replicated covariates, it can be stated that \(x_{iq} = u_{i} + \epsilon_{iq}\) where \(q=1,\dots, Q_{i}\) refers to the number of the replicated measurements for each \(x_{i}^{P=4}\) because the amount of replicate information can vary by each data point. Then the estimate of \(\sigma_{\epsilon}^2\) can be computed by
    \begin{equation*}
    \hat{\sigma_{\epsilon}^2} = \frac{\sum_{i=1}^{n} \sum_{q=1}^{Q_{i}}(x_{iq}-\bar{x_{i}})^2}{\sum_{i=1}^{n}(Q_{i}-1)}
    \end{equation*} In our study, we generate various replicates with bootstrapping technique to practice this idea, and \(\hat{\sigma_{\epsilon^{4}}^2}=134.56\) associated with the covariate \(x^{P=4}\) is obtained.      

\begin{figure}[!t]
\centering
\includegraphics[width=3.5in]{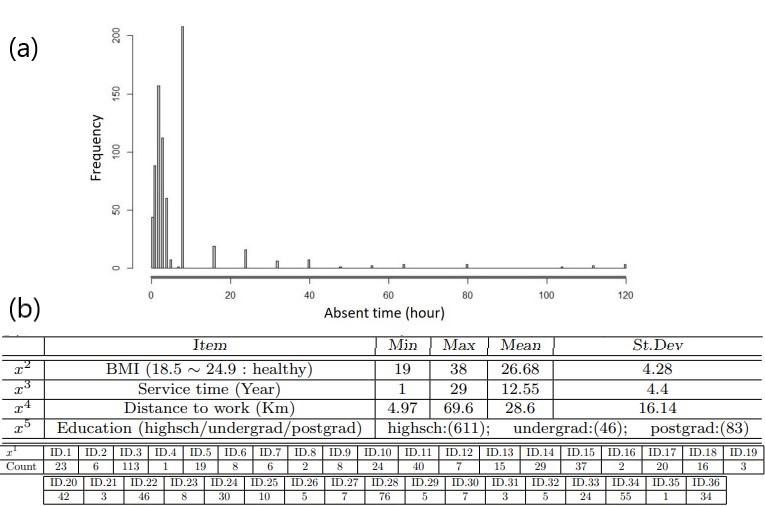}
\caption{A histogram of the outcome variable \(y\) absent count distribution (a) and summary descriptions of the selected covariates \(x^{P=2},x^{P=3},x^{P=4},x^{P=5}\) with the absence case counts by employee ID \(x^{P=1}\) (b). The continuous variable \(x^{P=4}\) - ``Distance to work" - is contaminated with measurement errors: \(\epsilon_{i} \sim \mathbf{N}(0, \sigma_{\epsilon}^2)\).}
\label{fig_sim}
\end{figure} 

Before fitting our hybrid Bayesian GLMM, we start by running the naive GLM with SIMEX for the sake of comparison. Bootstrapping is used because the overall distribution of the outcome variable shown in Figure (3) is not balanced and the sample size in certain clusters (i.e. employee ID) is not sufficient enough for direct statistical inference. The result of the naive model fitting with/without correction of \(x^{P=4}\) is provided in Figure (4) where we exclude the variables - \(x^{2}, x^{3}\) - due to the very large p-values 0.36 and 0.42 respectively. 
\begin{figure}[!t]
\centering
\includegraphics[width=3in]{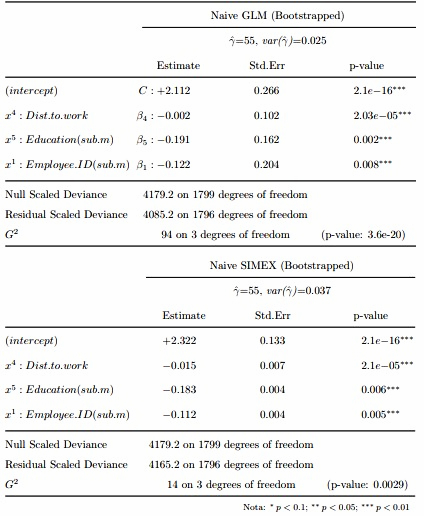}
\caption{The result of the naive \(\mathbf{NB}(\eta=\exp(X^{T}\boldsymbol{\underline{\beta}}), \; \gamma)\) regression analysis for absence hour count with error-prone `Dist.to.work' estimate \(\beta^{4}\). This is naive because the overdispersion and cohort properties in the outcome variable are not addressed. Note that only significant variables - \(x^{1},x^{4},x^{5}\) - are presented.}
\label{fig_sim}
\end{figure} 
From this model fitting results in Figure (4), it is clear to see the overall impact of overdispersion in \(y\) and the measurement error in \(x^{4}\) on the overall biased estimation. Although bootstrapped samples slightly reduce standard errors in the estimates, the SIMEX dramatically increases the overall residual deviance by correcting errors, which represents the improvement of the model fit.

\subsection{Implementation Results and Model Comparison}
The prior values for the regression coefficients \(\underline{\boldsymbol{\beta}}_{j}\) are initialized with SIMEX while the initial value for the dispersion term \(\gamma_{j}\) is estimated with the Method of Moments~\cite{al2010estimating}. As for the hyperparameters: \(\nu, \tau, s, t, u, v\) to estimate the hyper-prior: \(\underline{\mu}_{j},\Lambda_{j},k_{j},a_{j}\), we consider a non-informative prior approach as follows:
\begin{equation*}
\begin{split}
    &\nu=(\alpha+1)+(p+1)+1 \\ &\tau=(\alpha+1)\Sigma_{E} \\
    &s=0.001, \;t=1, \; u=1,\; v=1
\end{split}
\end{equation*} where \(\alpha\) = 0.01 tunes the prior knowledge as important as 1 observation in 100 samples, and \(\Sigma_{E}\) denotes the variance matrix chosen by the researcher. Further details of the hyperparameter choice above can be found in Fink and Bousquet~\cite{fink1997compendium,bousquet2008diagnostics}. 

\begin{figure}[!t]
\centering
\includegraphics[width=3.5in]{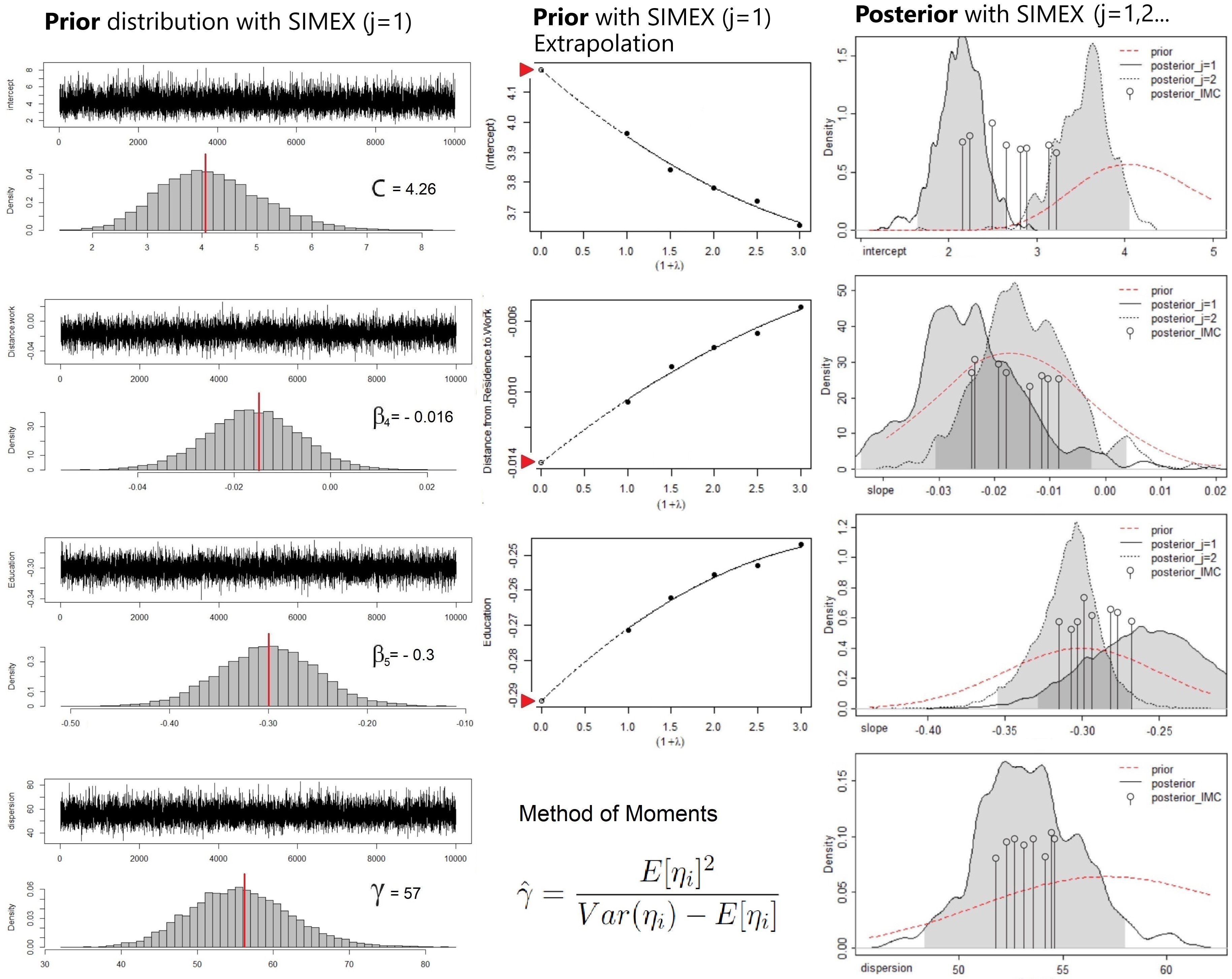}
\caption{Prior simulated from SIMEX, and the resulting posterior of \( C,\beta^{4}, \beta^{5}, \gamma \) computed with hierarchical Bayes. Note that only the first two models \(j=1,2\) are shown here due to page limitations.}
\label{fig_sim}
\end{figure} 

In order to implement our hierarchical Bayesian GLMM, we consider the regression equation with only the significant predictors - \(x^{1}, x^{4}, x^{5}\) -, that is given by
\begin{equation*}
\begin{aligned}
    &E[y_{j}|x^{1}_{j=1\sim36}] \\
    &= \mathbf{e}^{(intercept_{j} + \beta_{j}^{4}x_{j}^{4} + \beta_{j}^{5}I(x_{j}^{5}=a) + \beta_{j}^{5}I(x_{j}^{5}=b) + \beta_{j}^{5}I(x_{j}^{5}=c))}  
\end{aligned}
\end{equation*} where \(x^{5}\) has three levels - \(a,b,c\) -, and \(x^{1}_{j=1\sim36}\) contains 36 employees' IDs to examine the sources of the unobservable individual differences (i.e. overdispersion) in the outcome \(y\). For the sake of investigating the individual differences, we allow the 36 different regressions corresponding to each employee ID. For each individual model of \(j\), a training set of response and covariates pair \((y,X)\) with \(n\) = 50 records (i.e. total = 50 x 36) is constructed to build our model then an additional set of response and covariates pair \((y',X')\) with \(m\) = 100 (i.e. total = 100 x 36) records are fitted to the model to evaluate their performances. We perform 100 SIMEX simulations with the 1800 records. The series of plots given in Figure (5) succinctly illustrate the prior simulation and its posterior computation results - \(\boldsymbol{\underline{\beta}}, C, \gamma\). The plots show a bulk of different probability distributions from where the hierarchical Bayes algorithm simulates the dispersion parameter \(\gamma\) under a certain probability value and based on this simulation result, the posteriors \(\boldsymbol{\underline{\beta}}, C\) are re-sampled accordingly. The Durbin-Watson test for autocorrelation reveals p-values less than the level of significance - 0.05; hence, we fail to reject the null hypothesis of zero autocorrelation, indicating the chains are converged to the target distribution. 
\begin{figure}[!t]
\centering
\includegraphics[width=2.5in]{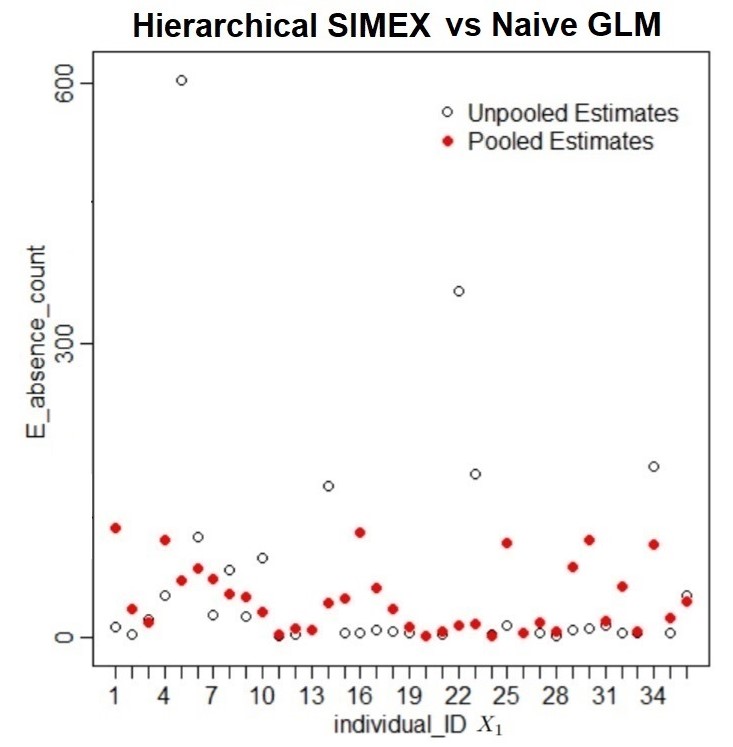}
\caption{Out-of-sample results of the expected absence hour counts \(E[y]_{j=1\sim36}\) of 36 individual employees. The estimates shown by the open points are produced from the naive GLM, and the other red points are the estimates produced from our hierarchical Bayes GLMM integrated with SIMEX.}
\label{fig_sim}
\end{figure} 

In order to see if our approach leads to substantial improvement of model fit and reduction of bias compared to the conventional naive approaches, we first refer to the plot in Figure (6) which illustrates the out-of-sample prediction results on the employees' expected absence hour counts \(E[y]\).  At a first glance, it is not clear how the adverse impact of the overdispersion and measurement error is addressed, but one can observe that the overall variation of group mean estimates is reduced. If the distinct values of the expected absence hour count for each individual can represent the identity of each cohort residing in the data, the implication of the reduced variation is the reduced number of cohorts via the estimation regularization. Thus, the plot in Figure(6) is a good sign as it shows that our hybrid model captures certain information running across different individuals, which reduces the bias problem triggered by the overdispersed outcome and mismeasured covariate. 

In Figure (7), the out-of-sample prediction performances of our hybrid Bayesian GLMM are compared with conventional models such as naive GLM and SIMEX model. It is apparent from the box plots that our model (i.e. hierarchical Bayes + SIMEX) leads to more improvement in model fit than the case with SIMEX only. The measurement error correction with SIMEX clearly abates the bias and improves model fit, but the hierarchical Bayes brings a more significant bias reduction, exhibiting dramatically smaller deviances and a narrower range of variance, compared with the plain SIMEX and GLM models. This observation becomes even more clear when we use advanced metric such as WAIC that accounts for the risk of overfitting. 
\begin{figure}[!t]
\centering
\includegraphics[width=3.3in]{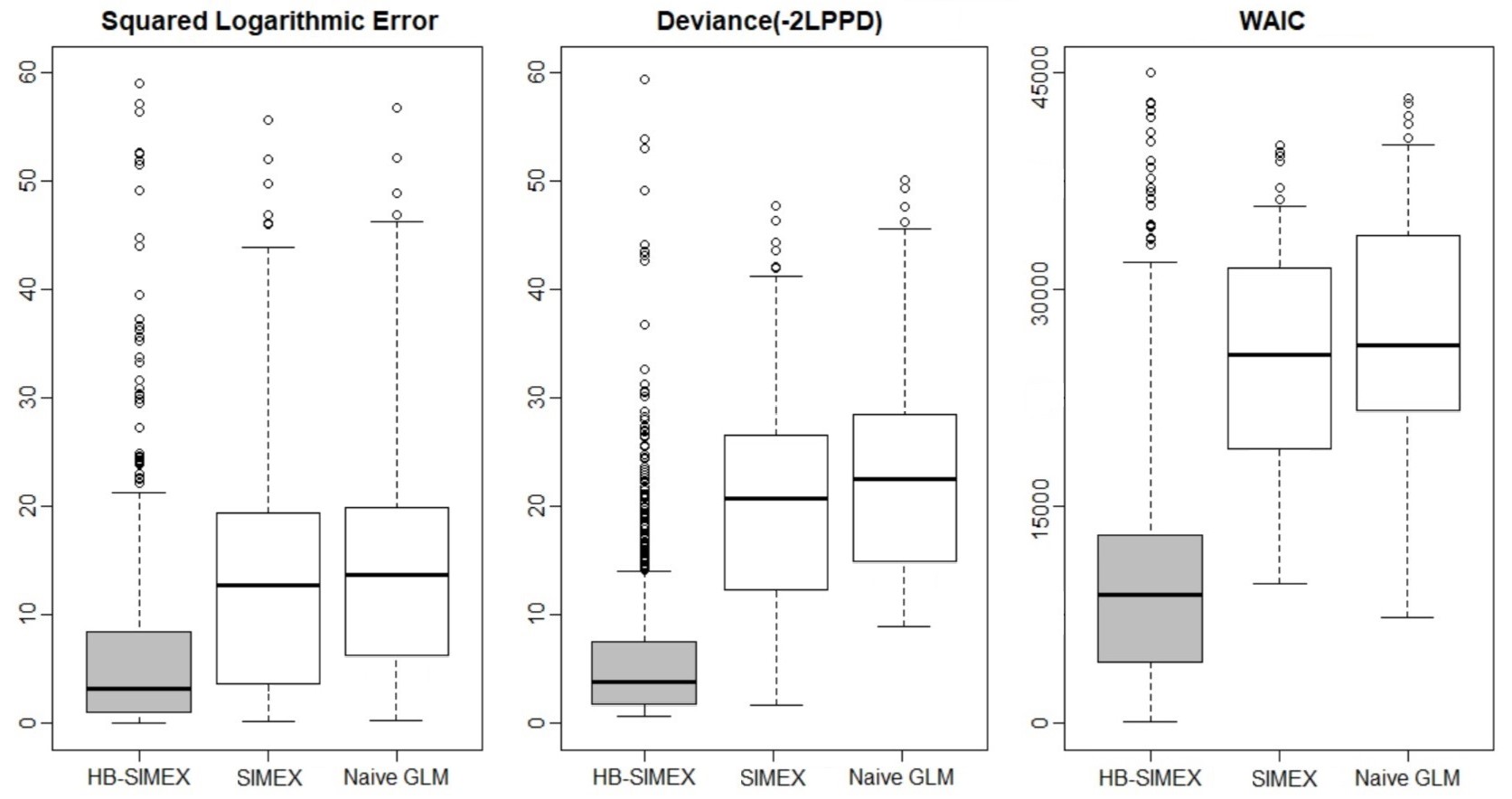}
\caption{Boxplots of three different model- evaluation metrics - Mean Squared Logarithmic Error, Scaled Deviance, and WAIC - built upon in-sample and out-of-sample results of the models - 1) Hierarchical Bayesian GLMM based on SIMEX, 2) GLM based on SIMEX, 3) Naive GLM.}
\label{fig_sim}
\end{figure} 

Figure (8) compares the two hierarchical Bayesian GLMMs with/without SIMEX. The first two plots on the top show the amount of the correlation between the varying slope \(\beta_{j}^{4}\) and varying intercept \(C_{j}\)~\footnote{ Dougherty~\cite{dougherty2011introduction} points out that the investigation of the correlation between the intercept and slope sometimes can help us unfold the structure of complex data. This is because it gives an idea of the degree to which the variation in ``cohort mean" (varying intercepts) is related to the variation in the effect of the ``cohort mean" on the outcome (varying slopes). Therefore, the variability in the unobservable relationships could be explained by differing cohorts.} learned by each hierarchical Bayesian model, and both land on reasonably high and negative values. The other two scatter plots on the bottom display the 36 estimate pairs of \(C, \beta^{4}\) shrunk towards the global mean, keeping the variability between the cohorts intact. This is because the hierarchical Bayesian GLMM simultaneously estimates both an ``intercept for each cohort" and the ``variation among cohorts" by modelling their covariance. What is important is that although both hierarchical Bayesian models discover the unobservable information that can redefine the relationship between each cohort, the model integrated with SIMEX captures much more of the information as a result of the error correction in the covariate. This interpretation is consistent with the box plots for the model fit comparison in Figure (7) as the SIMEX-based hierarchical model learning this correlation much deeper can model the data better by reducing the bias and improving the model fit. 

From the perspective of the insurer's risk classification, the properties of the insured class become much clearer. The high, negative correlation between the varying slope and intercept suggests that the employees who are absent more often than others on average are likely to show high sensitivity to the longer \emph{distance-to-work}: \(x^{4}\). On the other hand, those with a small number of absence hour counts tend to be more affected by other sources (i.e. covariates) than the travel distance \(x^{4}\); therefore, it is safe to say that 36 employee behaviors could differ in terms of how \emph{distance-to-work} affects their absence hour count. 
\begin{figure}[!t]
\centering
\includegraphics[width=3.7in]{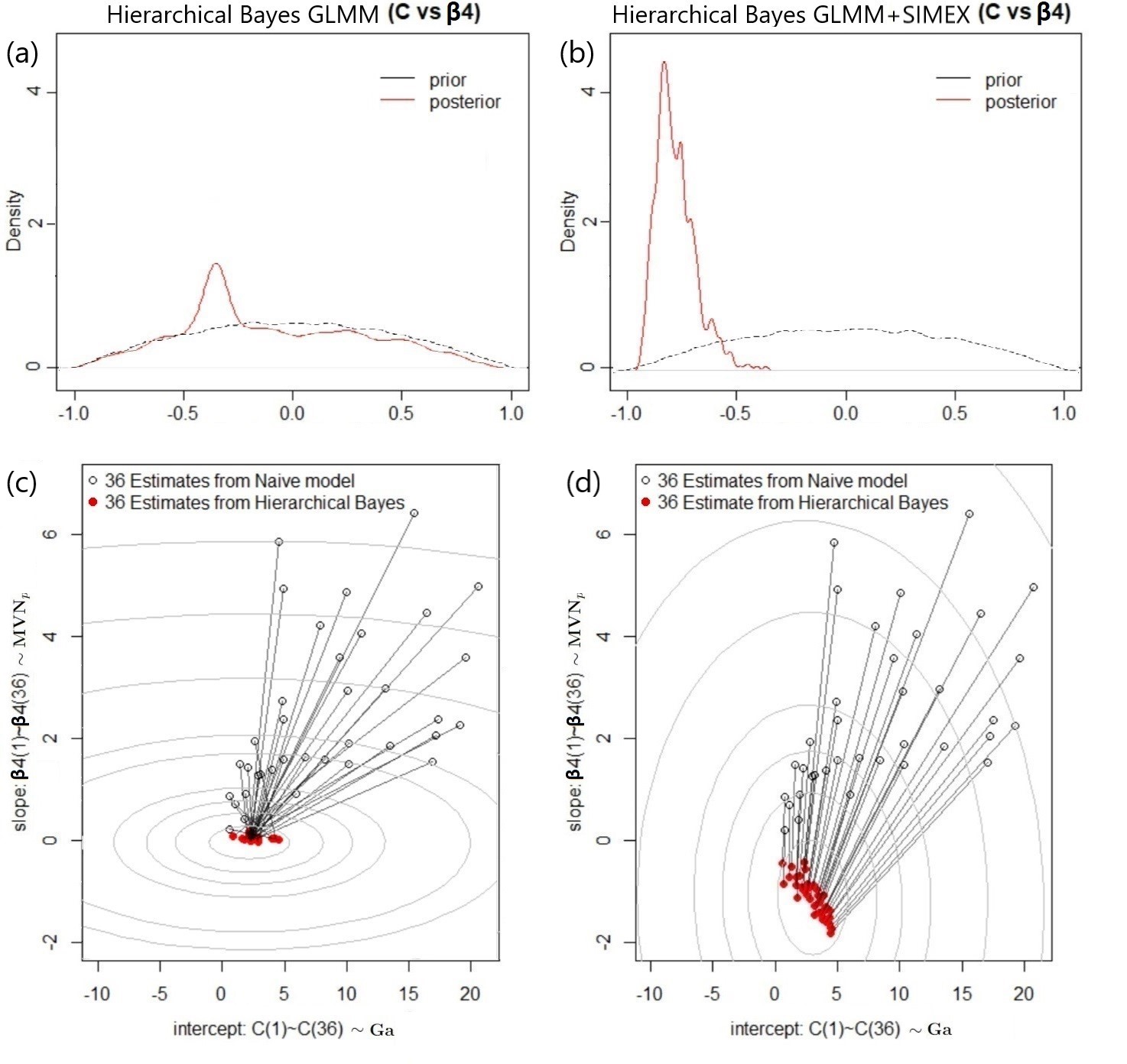}
\caption{Posterior density plots of the correlation term between the varying intercept \(C_{j}\) and varying slope \(\beta_{j}^{4}\) (a),(b). The scatter plots illustrating their subsequent shrinkage in the 36 estimates \(C_{j=1\sim36}, \beta^{4}_{j=1\sim36}\) from our hyerarchical models (c),(d). The estimates are produced from the Hierarchical GLMM without SIMEX (a),(c) and that with SIMEX (b,d). The contours depict the probability density from where the estimates are obtained.}
\label{fig_sim}
\end{figure}

\section{Discussion}
In this paper, we have extended a classical Hierarchical Bayesian GLMM by including an error correction method - SIMEX - to address overdispersed count-based outcome and mismeasured covariates at the same time. To our knowledge, the approach of combining a hierarchical Bayes with SIMEX has not previously been discussed in the literature. The result suggests that the hierarchical Bayesian GLMM + SIMEX outperforms the plain GLM+SIMEX or classical hierarchical Bayesian GLMM when proper prior knowledge of the error variance is available and the sample size is small in particular.

There are several limitations to this study. Above all, we need to reinforce the theoretical foundation in terms of 1) specification of the measurement error settings, 2) optimization of the hyperparameters, and 3) development of the proper extrapolation curve for SIMEX. 
\begin{enumerate}
    \item Although in this study, we assume a scenario of simple measurement errors that is independent of the other covariates, this is not always the case. More often than not, each individual cohort or covariate can influence the properties of the corresponding measurement error, which can make SIMEX obsolete since the distributions of the errors become inaccessible. If this is the case, we might have to consider other alternatives. \\
    
    \item It is difficult to choose the optimal hyperparameters for MCMC sampling mentioned in Section 2. There are a number of approaches we can consider such as ``data agreement criterion", ``empirical Bayes and marginal likelihood"~\cite{bousquet2008diagnostics}, etc. However, further research is necessary to explore how they interact with the estimation results. \\
    
    \item The exact parametric form of extrapolation function (describing the relation between degrees of error and the corresponding estimation results for the main parameters) for SIMEX is not available in most cases because it requires perfect true covariates to discover the true form. In our study, a popular quadratic extrapolation function is used, but we have to point out that it is just useful to obtain a good approximation. \\
\end{enumerate}

Lastly, given that our focus is on insurance applications, we should consider the extension of the predictive capacity of the resulting claim count distributions with other external information such as benefit claim amount or loss, etc. Some absence cases do not necessarily turn into benefit claims as the insurance company can refuse the claims when the cases do not meet their criteria. By matching the absence count prediction with the actual loss, a company can make a better interpretation of the distribution tails, and extreme values, or investigate a conversion rate that would be directly translated into future company expenses.


%




\section*{Disclosure statement}
The authors declare no conflict of interest.

\ifCLASSOPTIONcaptionsoff
  \newpage
\fi



%

\bibliographystyle{unsrt}
\bibliography{bibliography.bib} 

\end{document}